# Likelihood Analysis of Repeating in the BATSE Catalogue

Jean M. Quashnock[1]

*Department of Astronomy and Astrophysics*
*University of Chicago, Chicago, Illinois 60637*

I describe a new likelihood technique, based on counts-in-cells statistics, that I use to analyze repeating in the BATSE 1B and 2B catalogues. Using the 1B data, I find that repeating is preferred over non-repeating by 4.3:1 odds, with a well-defined peak at 5-6 repetitions per source. I find that the post-1B data are consistent with the repeating model inferred from the 1B data, after taking into account the lower fraction of bursts with well-determined positions. Combining the two data sets, I find that the odds favoring repeating over non-repeating are almost unaffected at 4:1, with a narrower peak at 5 repetitions per source. I conclude that the data sets are consistent both with each other and with repeating, and that for these data sets the odds favor repeating.

## INTRODUCTION

By studying clustering in the angular distribution of bursts, Quashnock and Lamb (1,2) found evidence that $\gamma$-ray burst sources repeat. Using a nearest neighbour analysis of the BATSE 1B catalogue (3), we found a significant excess of neighbours on angular scales less than $5°$. Since this is less than the median positional error of bursts in the catalogue, we concluded that burst sources repeat multiple times, on a time-scale of months. Wang and Lingenfelter (4) presented evidence that five particular bursts arise from a single repeating source: They also found evidence of repeating by studying spatial and temporal correlations in the 1B data (5). Using a particular model of repeating, Strohmayer, Fenimore and Miralles (6) also found evidence for repeating.

Clearly it was paramount to extend these analyses to the BATSE 2B catalogue (7). Unfortunately, the failure of the tape recorders on board the *Compton* Observatory led to a decrease of $\sim 1/3$ in the fraction of bursts with well-determined (so-called "non-MAXBC") positions in the new (or 2B-1B) data. While two-point angular correlation function and nearest neighbour analyses of the 2B catalogue failed to confirm repeating (8), this is expected given the positional errors of order 7° and the aforementioned drop in exposure (5). This is particularly the case if the typical number of observed repetitions from each source is small (9). Assuming that the bursts in the 1B catalogue form a fair sample, Lamb and Quashnock (10) simulated what would be expected in the 2B-1B and 2B catalogues. We found that repeating is not detectable, due to the lower fraction of bursts with non-MAXBC positions.

---

[1] *Compton* GRO Fellow – NASA grant GRO/PDP 93-08.

If burst sources do repeat, this would significantly constrain the range of allowed models and favor a Galactic origin. Given these implications and the controversy surrounding the issue of repeating, it is clearly imperative to use the most powerful and sensitive techniques to test the repeating hypothesis – in particular because of the drop of exposure in the 2B catalogue. Likelihood methods (11) are known to be the most sensitive and give the best possible determination of model parameters.

Here I describe a new likelihood method based on counts-in-cells that I have developed and used to analyze the clustering of gamma-ray bursts in the BATSE 1B and 2B catalogues (9). It allows us to test various repeating models parametrized by the number $N_r$ of repeating sources and the number $N_{rep}$ of bursts emitted by each repeating source, and to include the important effects of exposure and positional errors.

## COUNTS-IN-CELLS LIKELIHOOD

Let $N_{cell}$ be a large number of cells, each of fixed solid angle size $\Omega$, each centered on a random position on the sky. Let $C_N$ to be the number of these cells having $N$ bursts in them, where $N = 0, 1, 2, \ldots$. I then define the observed counts-in-cells distribution,

$$P_N \equiv C_N/N_{cell}, \qquad (1)$$

which is the probability that a randomly chosen cell of size $\Omega$ has $N$ bursts in it.

The counts-in-cells distribution contains information about clustering on scales comparable to the angular size of the cell. Indeed, the expected distribution $Q_N$ is Poisson when the bursts are uniformly distributed on the sky:

$$Q_N = \frac{1}{N!} e^{-\mathcal{N}\Omega} (\mathcal{N}\Omega)^N, \qquad (2)$$

where $\mathcal{N}$ is the number density of bursts.

I have calculated the expected counts-in-cell distribution $Q_N$ for a repeating model specified by the parameters $N_r$ and $N_{rep}$, where $N_r$ is the number of repeating sources each of which bursts exactly $N_{rep}$ times. By definition $N_{rep} \geq 2$. Note that this is the *actual* number of repetitions per source, not the *observed* number, which is on average much smaller because the BATSE sky exposure $\epsilon$ is significantly smaller than unity [$\epsilon = 0.34$ for the 1B catalogue and only 0.26 for the 2B-1B (non-MAXBC) catalogue]. I include both the effect of finite positional accuracy ($\theta_{\rm err} = 6.8°$ is the median error) and of sky exposure when calculating the expected counts-in-cells distribution in the repeating model. I also allow for an integer number $N_{\rm nr}$ of background sources that only burst once.

Once I calculate the expected counts-in-cells distribution $Q_N$ for a given set of parameters of the *model*, I ask how likely is the observed distribution $P_N$ for the *data*, given such a distribution. This likelihood $\mathcal{L}$ is given by the following (9):

$$\log \mathcal{L} = N_{\rm cell} \sum_N P_N \log Q_N + {\rm const.}. \qquad (3)$$

I then use Bayes' Theorem to interpret the likelihood in terms of a probability distribution for the integer values of the model parameters; namely the number of repeating sources $N_{\rm r}$ and the number of repetitions $N_{\rm rep}$ for each repeating source.

I have analyzed the BATSE 1B and 2B catalogues using this counts-in-cells technique and a cell size of 5°. Figure 1 (left panel) shows the 1-, 2- and 3-$\sigma$ contours in the ($N_{\rm rep}, N_{\rm r}$)-plane. The cross marks the maximum likelihood location. While the credible interval for the number of repeating sources is broad, that for the number of repetitions is considerably narrower. This is shown in Figure 1 (right panel), where I have marginalized over the number of repeating sources. There is a well-defined peak at 5 or 6 repetitions per source. The repeating model is favored over the non-repeating model by odds of 4.3:1.

Figure 2 (left panel) shows the results of the same analysis for the 2B-1B (non-MAXBC) catalogue. Note that the contours in the ($N_{\rm rep}, N_{\rm r}$)-plane are now much larger and extend down to smaller numbers of repeating sources and repetitions per source. Indeed, the odds of repeating versus non-repeating have fallen to 0.85:1 (basically equal odds). Nevertheless, the 1-$\sigma$ credible regions in Figures 1 and 2 largely overlap, and the maximum likelihood locations are quite close. Figure 2 (right panel) again shows the probability distribution for the number of repetitions per source. While the peak has now fallen to 2 repetitions, the probability of 5 repetitions is almost as large.

I have also combined the two data sets, and find the results shown in Figure 3. Interestingly, the contours of the 1-, 2-, and 3-$\sigma$ credible regions for the 2B catalogue (left panel) are actually smaller than for the 1B. This is evident in the right panel, which shows a well-defined peak in probability at 5 repetitions per source. Combining the two data sets, I find that the odds favoring repeating over non-repeating are almost unaffected, at 4:1, relative to that found for the 1B.

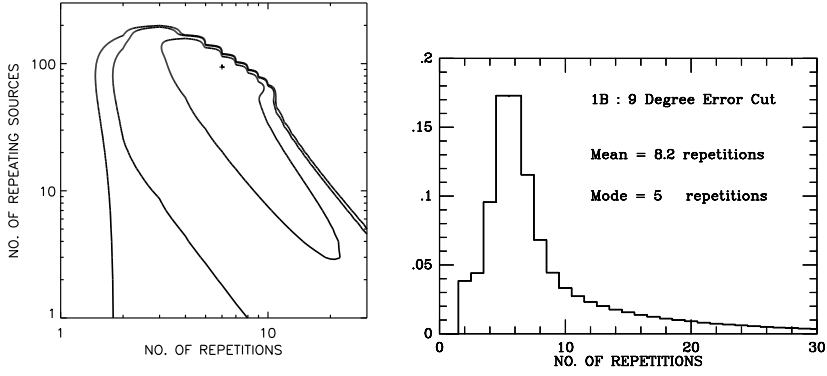

**FIG. 1.** (left panel) 1-, 2- and 3-$\sigma$ contours in the $(N_{\rm rep}, N_{\rm r})$-plane from analysis of the 1B catalogue. (right panel) Probability distribution of $N_{\rm rep}$ from the same analysis.

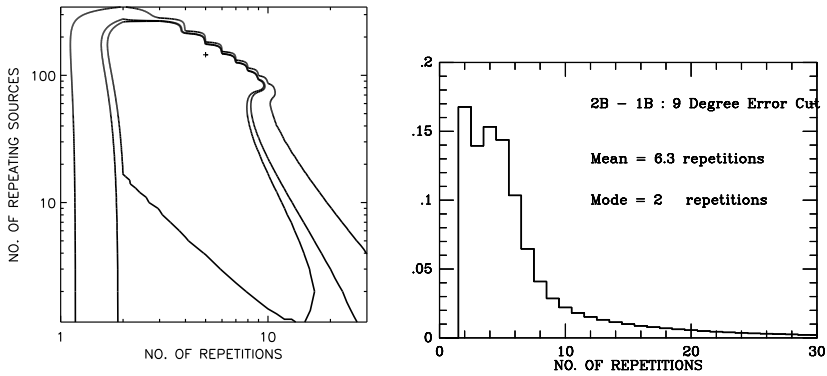

**FIG. 2.** Same as Figure 1, from analysis of the 2B-1B (non-MAXBC) catalogue.

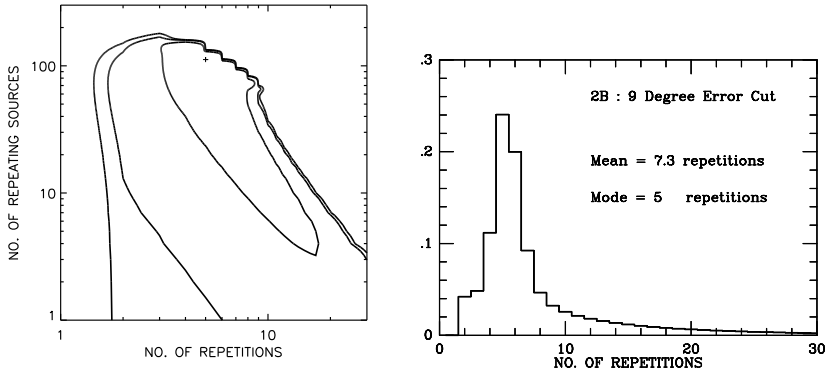

**FIG. 3.** Same as Figure 1, from analysis of the combined catalogues.

## CONCLUSIONS

Considering all of these results, I conclude that the first and second data sets are consistent both with each other and with repeating, and that for the combined data sets the odds still favor repeating. Thus the 2B-1B data alone neither confirm nor refute the repeating hypothesis, nor can they given the drop in exposure relative to the 1B catalogue (5). The 3B-2B catalogue containing 570 new bursts is not expected to suffer from this, and it should offer a fair test of the repeating hypothesis.

## REFERENCES

1. J. M. Quashnock and D. Q. Lamb, MNRAS **265**, L59 (1993).
2. J. M. Quashnock and D. Q. Lamb, *Gamma-Ray Bursts*: AIP Conference Proceedings 307, p. 107 (1994).
3. G. J. Fishman *et al.*, Ap. J. Suppl. **92**, 229 (1994).
4. V. C. Wang and R. E. Lingenfelter, Ap. J. **416**, L13 (1993).
5. V. C. Wang and R. E. Lingenfelter, Ap. J. **441**, 747 (1995).
6. T. E. Strohmayer, E. E. Fenimore and J. A. Miralles, Ap. J. **432**, 665 (1994).
7. C. A. Meegan *et al.*, electronic catalogue; grossc.gsfc.nasa.gov, username gronews, *Basic Information Table*, (1994).
8. C. A. Meegan *et al.*, Ap. J. Lett., (1995) (in press).
9. J. M. Quashnock, *Proceedings of the 29th ESLAB Symposium*, Astrophys. Sp. Sci., (1995) (in press).
10. D. Q. Lamb and J. M. Quashnock, *Proceedings of the 29th ESLAB Symposium*, Astrophys. Sp. Sci., (1995) (in press).
11. T. J. Loredo, *Statistical Challenges in Modern Astrophysics*, E. Feigelson and G. Babu, eds., (New York: Springer-Verlag), p.275 (1992).